\documentclass[11pt,a4paper]{article}
\usepackage{jheppub}
\usepackage{amsmath,amssymb,theorem,epsfig,url,psfrag,eepic,mathtools,mathrsfs,amsbsy,dsfont}
\usepackage{indentfirst}
\usepackage{graphicx}
\usepackage{tikz,pgfplots}
\pgfplotsset{compat=1.17}
\begin{document}
\title{\textbf{On loop corrections to integrable $2D$\\ sigma model backgrounds}\vspace*{.3cm}}
\date{}
\author[a,b]{Mikhail Alfimov}
\author[c,d]{and Alexey Litvinov}
\affiliation[a]{HSE University, 6 Usacheva str., Moscow 119048, Russia}
\affiliation[b]{P.N. Lebedev Physical Institute of the Russian Academy of Sciences, 53 Leninskiy pr., Moscow 119991, Russia}
\affiliation[c]{Landau Institute for Theoretical Physics, 1A Akademika Semenova av.,  Chernogolovka 142432, Russia}
\affiliation[d]{Center for Advanced Studies, Skolkovo Institute of Science and Technology, 1 Nobel str., Moscow 143026, Russia}
\emailAdd{malfimov@hse.ru} 
\emailAdd{litvinov@itp.ac.ru}
\abstract{
We study regularization scheme dependence of $\beta$-function for sigma models with two-dimensional target space. Working within four-loop approximation, we conjecture the scheme in which the $\beta$-function retains only two tensor structures up to certain terms containing $\zeta_3$. Using this scheme, we provide explicit solutions to RG flow equation  corresponding to Yang-Baxter- and $\lambda$-deformed $SU(2)/U(1)$ sigma models, for which these terms disappear.
}
\maketitle
\section{Introduction}

$\beta$-functions govern scale dependence of coupling constants in Quantum Field Theory. In  the textbook example of $\varphi^4$ theory in $4$ dimensions the coupling constant $g$ runs at two loops as
\begin{equation}\label{phi4-beta-function}
    \dot{g}=-\beta(g)\,, \quad \beta(g)=3g^2-\frac{17g^3}{3}+\ldots
\end{equation}
The terms depicted by $\dots$ in \eqref{phi4-beta-function} correspond to higher loop corrections. They have been computed (see \cite{Kompaniets:2016hct}) up to six loops (i.e. to the order $g^7$) using dimensional regularization in the minimal subtraction scheme \cite{tHooft:1972tcz,tHooft:1973mfk}. Higher loop coefficients in \eqref{phi4-beta-function} are scheme dependent. Change of regularization scheme corresponds to the ``coordinate change in the space of couplings''
\begin{equation}
  g\rightarrow\tilde{g}(g)=g+\xi_1 g^2+\xi_2 g^3+\ldots\,,
\end{equation}
where the parameters $\xi_k$ are the parameters of the new scheme. The $\beta$-function transforms as a vector field
\begin{equation}
  \dot{\tilde{g}}=\frac{\partial\tilde{g}}{\partial g}\dot{g}\implies
  \tilde{\beta}(\tilde{g})=\left(\frac{\partial \tilde{g}(g)}{\partial g}\right)^{-1}\beta(\tilde{g}(g))\,.
\end{equation}
For the $\beta$-function, which starts from $g^2$ (corresponding to classically marginal coupling constants) one has
\begin{equation}
  \beta(g)=A_1g^2+A_2g^3+A_3g^4+\ldots\implies
  \tilde{\beta}(\tilde{g})=A_1\tilde{g}^2+A_2\tilde{g}^3+(A_3+A_2\xi_1+A_1(\xi_1^2-\xi_2))\tilde{g}^4+\ldots
\end{equation}
Tuning the scheme parameters $\xi_k$, one can transform the $\beta$-function to some \emph{normal form}. In particular, one can define the scheme in which the $\beta$-function is two-loop exact.  

In case one considers theories with several coupling constants the situation becomes more involved. If $(g_1,\dots,g_n)$ correspond to local coordinates in the vicinity of some fixed point the RG (Renormalization Group) flow equations can be written in the form \cite{Wilson:1973jj}
\begin{equation}\label{generic-RG}
    \dot{g}_l=(D-\Delta_l)g_l+a^{ij}_l g_i g_j+b^{ijk}_l g_i g_j g_k+\ldots\,,
\end{equation}
where $\Delta_l$'s are dimensions of the operators corresponding to $g_l$. All the non-linear terms in \eqref{generic-RG} can be exterminated by the appropriate transformation of the coupling constants
\begin{equation}
    g_l\rightarrow g_l+\xi_1^{ij}g_ig_j+\xi_2^{ijk}g_ig_jg_k+\ldots\,,
\end{equation}
provided that the so called resonance terms are absent (see for example \cite{book-Ilyashenko}). In this case the RG flow equations \eqref{generic-RG} take very simple form: all $\beta$-functions are reduced to their linear parts (provided that all $\Delta_l\neq D$). This statement is known as (formal version of) Poincar\'e-Dulac theorem. 

One more example of using of  Poincar\'e-Dulac theorem besides the one considered above, with one marginal coupling, has been considered by Al. Zamolodchikov  in \cite{Zamolodchikov:1995xk} who applied PD theorem
to the sine-Gordon theory near the BKT transition point described by two classically marginal coupling constants $(g_{\parallel},g_{\perp})$. He showed that there is a scheme in which RG flow equations have the following exact form
\begin{equation}
    \dot{g}_{\parallel}=\frac{g_{\perp}^2}{1+\frac{g_{\parallel}}{2}}\,, \quad
    \dot{g}_{\perp}=\frac{g_{\parallel}g_{\perp}}{1+\frac{g_{\parallel}}{2}}\,.
\end{equation}
This scheme later has been successfully used in \cite{Doyon:2002ui,Lukyanov:2002fg} for calculation of correlation functions.

In these notes we study the question of scheme dependence for two-dimensional sigma models
\begin{equation}
  S[\boldsymbol{X}]=\frac{1}{4\pi}\int G_{ij}(\boldsymbol{X})\partial_{a}X^{i}\partial_{a}X^{j}\,d^{2}\boldsymbol{x}\,.
\end{equation}
Here $\boldsymbol{X}=(X^1,\dots,X^D)$ are treated as coordinates on some $D$-dimensional target manifold. The RG flow equations for the metric $G_{ij}(\boldsymbol{X})$ have the form
\begin{equation}\label{RG-flow-equation}
    \dot{G}_{ij}+\nabla_i V_j+\nabla_j V_i=-\beta_{ij}(G)\,,
\end{equation}
where $V_j$ correspond to field renormalization terms. The $\beta$-function $\beta_{ij}(G)$ admits covariant loop expansion
\begin{equation}\label{SM-beta-function-loop-expansion}
  \beta_{ij}(G)=\beta_{ij}^{(1)}(G)+\beta_{ij}^{(2)}(G)+\beta_{ij}^{(3)}(G)+\ldots\,,
\end{equation}
where the $L$-th loop order $\beta$-function coefficient $\beta_{ij}^{(L)}$ belongs to the finite dimensional space $\mathcal{T}_L$ of tensors with given scaling properties. Namely, if we assume that the metric is proportional to the inverse of the Planck constant
\begin{equation}\label{hbar-counting}
   G_{ij}\sim\hbar^{-1}\implies G^{ij}\sim\hbar,\;\Gamma^{k}_{ij}\sim \hbar^0,\;\nabla_i\sim\hbar^0,\;
   R_{ijk}^{\quad l}\sim\hbar^{0},\;R_{ij}\sim \hbar^0\quad\text{etc}.\,,
\end{equation}  
then the $L$-th loop order coefficient is a linear combination of tensors
\begin{equation}\label{beta_function_tensor_exp}
    \beta_{ij}^{(L)}=\sum_{k=1}^{\textrm{dim}\, \mathcal{T}_L}b_k^{(L)}\beta_{ij}^{k,L}(G)\,, \quad
    \beta_{ij}^{k,L}(G)\in\mathcal{T}_L\,, \quad \beta_{ij}^{k,L}(G)\sim\hbar^{L-1}\,.
\end{equation}
The coefficients $b_k^{(L)}$ were computed in the MS (Minimal Subtraction) scheme up to four loops. In particular, the one-loop $\beta$-function is proportional to the Ricci curvature \cite{Ecker:1971xko,Friedan:1980jm}
\begin{equation}\label{SM-beta-function-one-loop}
   \beta_{ij}^{(1)}=R_{ij}\,.
\end{equation}
Higher loop coefficients have been calculated: two-loop in \cite{Friedan:1980jm}
\begin{equation}\label{SM-beta-function-two-loop}
   \beta_{ij}^{(2)}=\frac{1}{2}R_{iklm}R_j^{\,\,klm}\,,
\end{equation}
three-loop in \cite{Graham:1987ep,Foakes:1987ij}
\begin{multline}\label{SM-beta-function-three-loop}
   \beta_{ij}^{(3)}=\frac{1}{8}\nabla_kR_{ilmn}\nabla^kR_j^{\,\,lmn}-\frac{1}{16}\nabla_iR_{klmn}\nabla_jR^{klmn}- \\
   -\frac{1}{2}R_{imnk}R_{jpq}^{\quad k}R^{mqnp}-\frac{3}{8}R_{iklj}R^{kmnp}R^{l}_{\,\,mnp}\,.
\end{multline}
Also the four-loop result has been obtained in \cite{Jack:1989vp}, but it is too long to be presented here\footnote{Below we present it in particular case of $D=2$ (see \eqref{beta-4}).}.

The higher loop coefficients $\beta_{ij}^{(L)}$ for $L>1$ are scheme dependent. If one stays within covariant approach, then different schemes are related to each other by covariant metric redefinitions
\begin{equation}\label{metric-redefinitions}
    G_{ij}\rightarrow G_{ij}+\sum_{k=0}^{\infty}G_{ij}^{(k)}\,,
\end{equation}
where $G_{ij}^{(k)}$ is of the order $\hbar^k$. Using this freedom and possibility to modify the vector $V_i$ in \eqref{RG-flow-equation}  one can try to bring the $\beta$-function to some normal form. This problem is highly non-trivial and unlikely to be solved in full generality since even the dimensions of the spaces $\mathcal{T}_k$ are not known in general. Moreover, the analysis performed at lower orders  \cite{Metsaev:1987zx} shows that there are tensor structures in the $\beta$-function beyond $L=1$ which are scheme invariant and it is highly likely that such structures persist at higher loops as well.  

Some hints to the solution of this problem can be obtained by considering integrable sigma models. In particular it has been noticed that some of the classically integrable sigma models are one-loop renormalizable \cite{Fateev:2019xuq} and that it might be a signal for quantum integrability. This question has been put forward in \cite{Hoare:2019ark} and later in \cite{Hoare:2019mcc,Hoare:2020fye,Levine:2021fof}, where it has been found that classical backgrounds require certain $\hbar$ corrections to persist renormalizability at higher loops. In particular, in the original paper \cite{Hoare:2019ark} the authors studied two integrable $2D$ backgrounds, namely the sausage model \cite{Fateev:1992tk} (in modern language the Yang-Baxter-deformed $SU(2)/U(1)$ sigma model) and the $\lambda$-deformed $SU(2)/U(1)$ sigma model \cite{Sfetsos:2013wia,Hollowood:2014rla}. In both cases they found $\hbar$ corrections to the classical backgrounds which provide solution to \eqref{RG-flow-equation} within three-loop approximation for the sausage model and within two-loop approximation for the $\lambda$ model. In our paper we improve the results of \cite{Hoare:2019ark}. Namely, working within four-loop approximation, we provide some universal scheme in which both backgrounds receive corrections of the order $\hbar^0$ only. Moreover, both $\hbar$-corrected backgrounds correctly reproduce the UV behaviour prescribed by dual Toda field theory corresponding to integrable perturbations of sine-Liouville CFT \cite{Fateev:2017mug}.

This paper is organized as follows. In the section \ref{2D-beta} we study in detail the four-loop $\beta$-function in $2D$ sigma models. We show that there exists a scheme in which the $\beta$-function contains only two tensor structures for certain $\hbar$-corrected versions of the sausage and $\lambda$-deformed $SU(2)/U(1)$ sigma models, which solve the four-loop RG flow equation. The curious property of these solutions is that they are ``two-loop exact'', i.e. contain only $\hbar^0$ corrections to the corresponding classical backgrounds.  In the section \ref{UV-dual} we consider the UV expansion of our solutions. In the section \ref{concl} we provide some remarks and directions for future work.
\section{\texorpdfstring{$\beta$}{beta}-function for sigma models with \texorpdfstring{$2D$}{2D} target space}\label{2D-beta}

Present section is devoted to the study of how the $\beta$-function changes under covariant metric redefinitions. This is done to find if there is a particular scheme in which it has a simple form, and then to search for the metrics, which solve the RG flow equation in this scheme.

In the case of two-dimensional target space $D=2$ all formulae are simplified since the Riemann tensor $R_{ijkl}$ can be algebraically expressed in terms of the scalar curvature $R$ and the metric
\begin{equation}\label{Riemann-2D}
    R_{ijkl}=\frac{R}{2}\left(G_{ik}G_{jl}-G_{il}G_{jk}\right)\,.
\end{equation}
Therefore all tensors in the $\beta$-function are expressed in terms of the metric, the scalar curvature $R$ and its covariant derivative. Substituting \eqref{Riemann-2D} into first four loop coefficients in the MS scheme (see \eqref{SM-beta-function-one-loop} and \eqref{SM-beta-function-three-loop}), one finds 
\begin{equation}\label{beta-123}
 \beta_{ij}^{(1)}=\frac{1}{2}RG_{ij}\,, \quad
 \beta_{ij}^{(2)}=\frac{1}{4}R^2G_{ij}\,, \quad
 \beta_{ij}^{(3)}=\left(\frac{5}{32}R^3+\frac{1}{16}\big(\nabla R\big)^2\right)G_{ij}-\frac{1}{16}\nabla_{i}R\nabla_{j}R
\end{equation}
and
\begin{multline}\label{beta-4}
  \beta_{ij}^{(4)}=\left(\frac{23}{192}R^4+\frac{2+\zeta(3)}{32}R^2\nabla^2R+\right. \\
  \left.+\frac{41+12\zeta(3)}{192}R\big(\nabla R\big)^2+\frac{1}{192}\big(\nabla^2R\big)^2+\frac{1}{192}\big(\nabla_k \nabla_l R\big)^2\right)G_{ij}- \\
  -\frac{\zeta(3)}{48}R^2\nabla_{i}\nabla_{j}R-\frac{25+8\zeta(3)}{192}R\nabla_{i}R\nabla_{j}R-\frac{1}{96}\big(\nabla^2R\big)\nabla_{i}\nabla_{j}R\,.
\end{multline}

Then the first three orders of the most general covariant metric redefinition \eqref{metric-redefinitions} take the form 
\begin{align}
  G_{ij}^{(0)} &= c_1 R G_{ij}\,, & \notag \\
  G_{ij}^{(1)} &= \left(c_2R^2+c_3\nabla^2R\right)G_{ij}+c_4\nabla_{i}\nabla_{j}R\,, & \notag \\
  G_{ij}^{(2)} &= \left(c_5R^3+c_6\big(\nabla R\big)^2+c_7R\nabla^2R+c_8\nabla^2\nabla^2R\right)G_{ij}+c_9\nabla_{i}R\nabla_{j}R+c_{10}R\nabla_{i}\nabla_{j}R+ \notag \\
  &+c_{11}\nabla_{i}\nabla_{j}\nabla^2R\,. \label{cov_red_metric}
\end{align}
In appendix \ref{schemes} we show that one can tune the scheme coefficients $c_p$ in such a way, that the $\beta$-function takes the form
\begin{multline}\label{all-loop-beta-function-first-4-loops}
\beta_{ij}=\left(\frac{R}{2}+\frac{R^2}{4}+\frac{3R^3}{16}+\frac{5R^4}{32}+\frac{2+\zeta_3}{64}\nabla^2\left(R^3+2R\nabla^2R-\frac{1}{2}\nabla^2R^2\right)+\dots\right)G_{ij}-\\-
\left(\frac{1}{16}+\frac{5R}{32}+\dots\right)\nabla_{i}R\nabla_{j}R+\ldots\,,
\end{multline}
where $\ldots$ correspond to terms coming from the $5$-th and higher loops and we omitted the tildes which denote the quantities in the new scheme. 

It is important to emphasise that the terms in the fourth-loop $\beta$-function coefficient proportional to $(2+\zeta_3)$ cannot be exterminated completely by the change of scheme. This phenomenon is well known and has been noticed first in the supersymmetric case \cite{Grisaru:1986px,Grisaru:1986dk}. Moreover there is some evidence that such terms persist at higher loop orders \cite{Grisaru:1986wj,Jack:1992ce}. On the other  hand, one can argue that physically acceptable backgrounds should not contain transcendental terms like $\zeta_n$. This is confirmed by the exact form of the cigar metric \cite{Dijkgraaf:1991ba} and by the all-loop metric for the sausage model found in \cite{Hoare:2019ark}. Both of them do not contain  $\zeta_n$ terms. Thus the presence of the $\zeta_n$ terms in the $\beta$-function and their absence in explicit solutions might seem puzzling.

We have found that the resolution of this puzzle in the fourth loop order lies in the fact that all metrics mentioned above (and also $\lambda$-deformed $SU(2)/U(1)$ metric) satisfy the following special relation
\begin{equation}\label{special-condition}
    \nabla^2\left(R^3+2R\nabla^2R-\frac{1}{2}\nabla^2R^2\right)=O(\hbar^{5})\,,
\end{equation}
and hence the terms in \eqref{all-loop-beta-function-first-4-loops} proportional to $(2+\zeta_3)$ can be dropped. We note that in this  case  only two tensor structures in the $\beta$-function are present, thus suggesting the general ansatz\footnote{We can not resist to mention that the sequence in  \eqref{all-loop-beta-function-first-4-loops} (with the terms $\sim(2+\zeta_3)$ dropped) can be easily extrapolated to 
\begin{equation}\label{all-loop-beta-function}\tag{*}
\beta_{ij}=\frac{RG_{ij}}{2(1-R)^{\frac{1}{2}}}-\frac{1}{16(1-R)^{\frac{5}{2}}}\nabla_{i}R\nabla_{j}R\,.
\end{equation}
It is possible that this ``exact'' expression for $\beta_{ij}$ as well as the generic ansatz \eqref{beta-generic-ansatz} could be broken by five-loop calculations (in particular by the presence of $\zeta_4$ terms), but checking this is a tedious task going beyond the scope of this paper. However, we will find that \eqref{all-loop-beta-function} is confirmed indirectly by the fact that $\hbar$-corrected backgrounds presented below (see \eqref{sausage-all-loop} and \eqref{lambda-deformed-metric-all-loop}) solve not only the four-loop RG flow equation, but also the ``exact'' equation as well. We will discuss the ``exact'' expression \eqref{all-loop-beta-function} in the section \ref{UV-dual}.}
\begin{equation}\label{beta-generic-ansatz}
    \beta_{ij}=A(R)G_{ij}+B(R)\nabla_iR\nabla_jR\,.
\end{equation}
We claim that the scheme in which the $\beta$-function has the form \eqref{all-loop-beta-function-first-4-loops} (with the terms $\sim(2+\zeta_3)$ dropped) is natural because both classically integrable backgrounds, $SU(2)/U(1)$ YB and $\lambda$-deformed models, admit simple $\hbar$ extension. We study them separately in the next subsections.
\subsection{Sausage model}

It is convenient to define the sausage model as inhomogeneous YB-deformed coset sigma model \cite{Klimcik:2008eq,Delduc:2013fga} (also known as $\eta$-deformed sigma model). The classical action of the YB-deformed $SU(2)/U(1)$ sigma model is defined by Drinfel’d-Jimbo solution $\mathcal{R}$ to the modified YB equation and has the form
\begin{equation}\label{Coset-action-deformed}
\mathcal{S}=\frac{\kappa}{4\pi\hbar}\int\textrm{Tr}\left(
\left(\mathbf{g}\partial_{+}\mathbf{g}^{-1}\right)^{(\textrm{c})}\,\frac{1}{1-i\kappa\mathcal{R}_{\mathbf{g}}\circ\mathrm{P}_{\textrm{c}}}\,
\left(\mathbf{g}\partial_{-}\mathbf{g}^{-1}\right)^{(\textrm{c})}\right) d^{2}x\,,
\end{equation}
where $\mathbf{g}\in SU(2)$, $\mathcal{R}_{\mathbf{g}}=\textrm{Ad}\, \mathbf{g}\circ\mathcal{R}\circ\textrm{Ad}\,\mathbf{g}^{-1}$ and $\mathrm{P}_{\textrm{c}}$ is the projector on the coset space.   Using the following parametrization of the coset elements
\begin{equation}
\mathbf{g}^{-1}=e^{\frac{i\chi}{2}\sigma_3}e^{\frac{i\theta}{2}\sigma_1}\,,
\end{equation}
where $\sigma_3$ corresponds to $U(1)$, we arrive to the sausage metric
\begin{equation}\label{sausage-1-loop}
    ds^2=\frac{2\kappa}{\hbar}\frac{d\theta^2+\cos^2\theta d\chi^2}{1-\kappa^2\sin^2\theta}\;.
\end{equation}
This theory is renormalizable at one loop \cite{Fateev:1992tk} provided that $\hbar$ does not run and $\kappa$ satisfies
\begin{equation}\label{kappa-one-loop-flow-equation}
    \dot{\kappa}=\frac{\hbar(\kappa^2-1)}{2}\,.
\end{equation}

The question of renormalizability of the sausage model at higher loops has been first considered in \cite{Hoare:2019ark}. It has been noticed in \cite{Hoare:2019ark} that renormalizability persists to higher loops subject to the addition of $\hbar$ corrections to the original metric \eqref{sausage-1-loop}. The authors in \cite{Hoare:2019ark} proposed some corrected metric and checked within the three loop approximation that there exists a scheme (different from ours \eqref{all-loop-beta-function-first-4-loops}) in which their metric solves the RG flow equation. We propose an alternative $\hbar$ completion of the one-loop metric \eqref{sausage-1-loop}
\begin{equation}\label{sausage-all-loop}
    ds^2=\frac{2\kappa}{\hbar}\frac{\left(1-\frac{\hbar\kappa\cos^2\theta}{1-\kappa^2\sin^2\theta}\right)d\theta^2+\cos^2\theta d\chi^2}{1-\kappa^2\sin^2\theta}\,,
\end{equation}
which solves the four-loop RG equation \eqref{RG-flow-equation} with the $\beta$-function \eqref{all-loop-beta-function-first-4-loops} provided that the coupling constant $\kappa$ satisfies the following flow equation
\begin{equation}\label{hbar-kappa-flow-equations-four-loops}
    \dot{\kappa}=\frac{\hbar(\kappa^2-1)}{2}\left(1+\frac{\nu}{2}\hbar
    +\frac{3\nu^2-4}{8}\hbar^2+\frac{5\nu^3-12\nu}{16}\hbar^3+\ldots\right)\,,
\end{equation}
and the vector field has the form
\begin{equation}\label{hbar-vector-field-equations-four-loops}
    \boldsymbol{V}=\hbar\left\{\frac{\kappa(\kappa^2-1)\sin2\theta}{4(1-\kappa^2\sin^2\theta)^2},
    \frac{\cos^2\theta}{1-\kappa^2\sin^2\theta}\right\}\left(1+\frac{\nu}{2}\hbar
    +\frac{3\nu^2-4}{8}\hbar^2+\frac{5\nu^3-12\nu}{16}\hbar^3+\ldots\right)\;,
\end{equation}
where $\nu=\kappa+\kappa^{-1}$.

We note that the metric \eqref{sausage-all-loop} is two-loop exact and only the flow equation \eqref{hbar-kappa-flow-equations-four-loops} and the vector field \eqref{hbar-vector-field-equations-four-loops} acquire higher loop corrections.
\subsection{\texorpdfstring{$\lambda$}{lambda}-deformed sigma model}

Now we present $\hbar$-corrected $\lambda$-deformed $SU(2)/U(1)$ metric. The classical ($1$-loop) metric is governed by the Lagrangian \cite{Sfetsos:2013wia,Hollowood:2014rla} 
\begin{equation}
    \mathcal{L}=\frac{1}{2\pi \hbar}\textrm{Tr}\left(-\frac{1}{2}\big(g^{-1}\partial g\big)\big(g^{-1}\bar{\partial}g\big)+J\bar{A}-A\bar{J}+g^{-1}Ag\bar{A}-A\bar{A}+\big(1-\lambda^{-1}\big)A\mathbb{P}\bar{A}
    \right)+\mathcal{L}_{\text{WZ}}\,,
\end{equation}
where $g\in SU(2)$
\begin{equation*}
   J=g^{-1}\cdot\partial g\,, \quad \bar{J}=\bar{\partial}g\cdot g^{-1} \quad \text{and} \quad
   \mathbb{P}=\mathbb{P}_{\frac{SU(2)}{U(1)}}\,,
\end{equation*}
$\mathcal{L}_{\text{WZ}}$ is the standard Wess-Zumino term, which is not important in the present case and $\lambda$ is the deformation parameter. Fixing the gauge symmetry as in \cite{Hoare:2019ark,Hoare:2019mcc} 
\begin{equation}
    g=e^{i\alpha\sigma_3}e^{i\beta\sigma_2} \quad \text{with} \quad \cos\alpha=\sqrt{p^2+q^2}\,, \quad \tan\beta=\frac{p}{q}\,,
\end{equation}
and integrating over the auxiliary fields  $(A,\bar{A})$, one arrives to the metric
\begin{equation}\label{lambda-deformed-metric-l-loop}
    ds^2=\frac{2}{\hbar}\frac{\kappa dp^2+\kappa^{-1}dq^2}{1-p^2-q^2}\,, \quad \text{where} \quad \kappa=\frac{1-\lambda}{1+\lambda}\,.
\end{equation}
This metric is one-loop renormalizable \cite{Itsios:2014lca,Appadu:2015nfa} with $\kappa$ running according to \eqref{kappa-one-loop-flow-equation} and the vector field given by
\begin{equation}
    V_p=\frac{p}{1-p^2-q^2}\,, \quad V_q=\frac{q}{1-p^2-q^2}\,.
\end{equation}

It has been demonstrated in \cite{Hoare:2019ark} that in order to ensure the two-loop renormalizability of the model one has to add a particular quantum correction coming from integration over the gauge field $(A,\bar{A})$
\begin{equation}
    ds^2\rightarrow ds^2-\frac{1}{2}\left[d\log(1-p^2-q^2)\right]^2=
    \frac{2}{\hbar}\left(\frac{\kappa dp^2+\kappa^{-1}dq^2}{1-p^2-q^2}-\hbar\frac{\big(pdp+qdq\big)^2}{\big(1-p^2-q^2\big)^2}\right)
\end{equation}
and properly modify the vector field.

We propose another $\hbar$ completion of \eqref{lambda-deformed-metric-l-loop} which is also two-loop exact similar to \eqref{sausage-all-loop}
\begin{equation}\label{lambda-deformed-metric-all-loop}
    ds^2=\frac{2}{\hbar}\left(\frac{\big(\kappa-\hbar\big)dp^2+\big(\kappa^{-1}-\hbar\big)dq^2}{1-p^2-q^2}-\hbar\frac{\big(pdp+qdq\big)^2}{\big(1-p^2-q^2\big)^2}\right)\,.
\end{equation}
This metric \eqref{lambda-deformed-metric-all-loop} solves the RG flow equation at four loops with the $\beta$-function \eqref{all-loop-beta-function-first-4-loops} provided that the coupling constant $\kappa$ satisfies \eqref{hbar-kappa-flow-equations-four-loops} and the vector field has the form
\begin{equation}\label{hbar_vector_field_equations_lambda_four_loops}
    V_p=\frac{p}{1-p^2-q^2}\left(1+V_p^{(1)}\hbar+V_p^{(2)}\hbar^2+V_p^{(3)}\hbar^3+\dots\right)\,, \quad
    V_q=\{p\rightarrow q,\kappa\rightarrow\kappa^{-1}\}\,,
\end{equation}
where the coefficients $V_{p}^{(k)}$ appear in the expansion of the function 
\begin{equation}
    \left(\frac{1-\hbar\kappa}{1-\hbar\kappa^{-1}}\right)^{\frac{1}{2}}\left(1-\frac{\hbar}{2\kappa}\frac{1-\left(\frac{1-\kappa^2}{1-\hbar\kappa}\right)q}{1-p^2-q^2}\right)=1+V_p^{(1)}\hbar+V_p^{(1)}\hbar+V_p^{(2)}\hbar^2+V_p^{(3)}\hbar^3+\ldots
\end{equation}

Having now two solutions \eqref{sausage-all-loop} and \eqref{lambda-deformed-metric-all-loop} it becomes possible to analyze their UV behaviour and check out how this behaviour correlates with the screening charges which define the sausage and $\lambda$-model. To achieve this we explain some curious observation concerning the form of the ``exact'' $\beta$-function in the next section.
\section{UV expansion}\label{UV-dual}
In the previous sections we managed to find two metrics \eqref{sausage-all-loop} and \eqref{lambda-deformed-metric-all-loop} which solve the RG flow equation at four loops with the $\beta$-function \eqref{all-loop-beta-function-first-4-loops} and vector fields \eqref{hbar-vector-field-equations-four-loops} and \eqref{hbar_vector_field_equations_lambda_four_loops} respectively and now would like to study the UV (or $t \rightarrow -\infty$) behaviour of these metrics.

As we already mentioned in the section \ref{2D-beta} the four-loop result for the $\beta$-function \eqref{all-loop-beta-function-first-4-loops} can be easily extrapolated to the following ``exact'' expression
\begin{equation}\label{all-loop-beta-function-concl}
\beta_{ij}=\frac{RG_{ij}}{2(1-R)^{\frac{1}{2}}}-\frac{1}{16(1-R)^{\frac{5}{2}}}\nabla_{i}R\nabla_{j}R\,.
\end{equation}
This form is remarkable in a sense that both solutions that we have found, namely \eqref{sausage-all-loop} and \eqref{lambda-deformed-metric-all-loop}, go through it, but with modified vector field and flow equations. More precisely,  \eqref{sausage-all-loop} and \eqref{lambda-deformed-metric-all-loop} solve the RG flow equation \eqref{RG-flow-equation} with the $\beta$-function given by \eqref{all-loop-beta-function-concl} provided that $\kappa$ flows according to the ``exact'' equation
\begin{equation}\label{hbar-kappa-flow-equations}
    \dot{\kappa}=\frac{\hbar(\kappa^2-1)}{2\left((1-\hbar\kappa)(1-\hbar\kappa^{-1})\right)^{\frac{1}{2}}}
\end{equation}
and the vector field has the form
\begin{equation}
    \boldsymbol{V}=\frac{\hbar}{\sqrt{(1-\hbar\kappa)(1-\hbar\kappa^{-1})}}\left\{\frac{\kappa(\kappa^2-1)\sin2\theta}{4(1-\kappa^2\sin^2\theta)^2},
    \frac{\cos^2\theta}{1-\kappa^2\sin^2\theta}\right\}\,,
\end{equation}
for the sausage metric \eqref{sausage-all-loop} and
\begin{equation}
    V_p=\frac{p\left(\frac{1-\hbar\kappa}{1-\hbar\kappa^{-1}}\right)^{\frac{1}{2}}}{1-p^2-q^2}\left(1-\frac{\hbar}{2\kappa}\frac{1-\left(\frac{1-\kappa^2}{1-\hbar\kappa}\right)q}{1-p^2-q^2}\right)\,, \quad V_q=\{p\leftrightarrow q,\kappa\rightarrow\kappa^{-1}\}
\end{equation}
for the $\lambda$-deformed one.

The UV stable solution of \eqref{hbar-kappa-flow-equations} admits the following UV expansion\footnote{We note that the differential equation \eqref{hbar-kappa-flow-equations} can be explicitly integrated 
\begin{equation}\tag{**}
    \left(\frac{1-\rho-\hbar}{1+\rho-\hbar}\right)^{1-\hbar}\left(\frac{1+\rho+\hbar}{1-\rho+\hbar}\right)^{1+\hbar}=e^{2\hbar(t-t_0)}\,,
\end{equation}
where $\rho\overset{\text{def}}{=}\sqrt{(1-\hbar\kappa)(1-\hbar\kappa^{-1})}$.
}
\begin{equation}\label{lambda_definition}
    \kappa=\frac{1-\lambda}{1+\lambda}\implies\lambda=e^{\frac{\hbar(t-t_0)}{1-\hbar}}+\frac{\hbar}{(1-\hbar)^2}
    e^{\frac{3\hbar(t-t_0)}{1-\hbar}}+\frac{\hbar(1+4\hbar+\hbar^2)}{(1-\hbar)^4}e^{\frac{5\hbar(t-t_0)}{1-\hbar}}+\ldots\,,
\end{equation}
which is qualitatively the same as the one-loop UV expansion \cite{Fateev:1992tk} with the only difference that the dimension of $\lambda$ is changed: $\hbar\rightarrow\frac{\hbar}{1-\hbar}$.

In order to study the UV behaviour more carefully, it is convenient to go to the other coordinate frame. Namely, for both models \eqref{sausage-all-loop} and \eqref{lambda-deformed-metric-all-loop} we perform the following change of variables (note that this change is complex)
\begin{align}
&\sin\theta=\kappa^{-1}\tanh\frac{x}{2}\,, & &\chi=\frac{y}{2}+\frac{i}{2}\log\left(1-\frac{1-\kappa^2}{1+\kappa^2}\cosh x\right) & &\text{for sausage model}\,, \notag \\
&p^2+q^2=e^{iy}\,, & &\frac{p^2-q^2}{p^2+q^2}=\cosh x & &\text{for $\lambda$ model}\,.
\end{align}
Then one has
\begin{equation}\label{all_loop_sausage_metric_exp}
ds^2_{\text{sausage}}=Adx^2+\left(A+\frac{1}{2}\right)dy^2+B\left(e^x(dx+idy)^2+e^{-x}(dx-idy)^2\right)
\end{equation}
and 
\begin{multline}\label{all_loop_lambda_metric_exp}
ds^2_{\lambda}=\frac{1}{1-e^{-iy}} \times \\
\times \left(Adx^2+\frac{1-\frac{A}{A+\frac{1}{2}}e^{-iy}}{1-e^{-iy}}
\left(A+\frac{1}{2}\right)dy^2+B\left(e^x(dx+idy)^2+e^{-x}(dx-idy)^2\right)\right)\,,
\end{multline}
where the parameters $A$ and $B$ have the form ($\lambda$ is defined by \eqref{lambda_definition})
\begin{equation}
    A=\frac{1+\kappa^2}{4\hbar\kappa}-\frac{1}{2}=\frac{1-\hbar}{2\hbar}+O(\lambda^2)\,, \quad B=\frac{1-\kappa^2}{8\hbar\kappa}=\frac{\lambda}{2\hbar}+O(\lambda^2)\,.
\end{equation}
In this form it is clear how these two models are related. Namely, one can read off the exponent $e^{-iy}$ in the metric \eqref{all_loop_lambda_metric_exp} by shifting $y$ by an infinite imaginary number (similar observation has been done in \cite{Hoare:2015gda}).

It is known from \cite{Litvinov:2018bou} that the theory \eqref{all_loop_sausage_metric_exp} in the UV limit can be understood as the free theory perturbed by some terms determined by certain exponential screening charges. In order to extract this form for this theory and try to do this for the $\lambda$ model \eqref{all_loop_lambda_metric_exp} it is convenient to rescale
\begin{equation}
    x\rightarrow\frac{x}{\sqrt{2A}}\,, \quad y\rightarrow\frac{y}{\sqrt{2A+1}}
\end{equation}
and then we obtain for the sausage sigma model
\begin{multline}\label{sausage_metric_rescaled}
    ds^2_{\text{sausage}}=\frac{1}{2}\left(dx^2+dy^2\right)+ \\
    +\frac{B}{2A}
    \left(e^{\frac{x}{\sqrt{2A}}}\left(dx+i\frac{\sqrt{2A}}{\sqrt{2A+1}}dy\right)^2+e^{-\frac{x}{\sqrt{2A}}}\left(dx-i\frac{\sqrt{2A}}{\sqrt{2A+1}}dy\right)^2\right)
\end{multline}
and for the $\lambda$-deformed sigma model
\begin{multline}\label{lambda_metric_rescaled}
    ds^2_{\lambda}=\frac{1}{2}\left(\frac{1}{1-e^{-i\frac{y}{\sqrt{2A+1}}}}dx^2+\frac{1-\frac{2A}{2A+1}e^{-i\frac{y}{\sqrt{2A+1}}}}{\left(1-e^{-i\frac{y}{\sqrt{2A+1}}}\right)^2}dy^2\right)+ \\
    +\frac{\frac{B}{2A}}{1-e^{-i\frac{y}{\sqrt{2A+1}}}}
    \left(e^{\frac{x}{\sqrt{2A}}}\left(dx+i\frac{\sqrt{2A}}{\sqrt{2A+1}}dy\right)^2+e^{-\frac{x}{\sqrt{2A}}}\left(dx-i\frac{\sqrt{2A}}{\sqrt{2A+1}}dy\right)^2\right)\,.
\end{multline}
Coefficients $A$ and $B$ have the following $t \rightarrow -\infty$ expansion
\begin{equation}\label{AB_UV_expansion}
    A=\frac{1-\hbar}{2\hbar}+\mathcal{O}\left(e^{\frac{2\hbar(t-t_0)}{1-\hbar}}\right)\,, \quad B=\frac{1-\hbar}{2\hbar}\hbar^{\frac{\hbar}{1-\hbar}}e^{\frac{\hbar(t-t_0)}{1-\hbar}}+\mathcal{O}\left(e^{\frac{3\hbar(t-t_0)}{1-\hbar}}\right)\,.
\end{equation}
Substituting the expansions \eqref{AB_UV_expansion} into the metrics \eqref{sausage_metric_rescaled} and \eqref{lambda_metric_rescaled}, we are able to extract the UV expansion of these metrics. By supplementing it with a new parameter together with an additional imaginary shift of $y$ coordinate
\begin{equation}\label{b_def_shift}
    b\overset{\text{def}}{=}\sqrt{\frac{1-\hbar}{\hbar}}\,, \quad y \rightarrow y+\frac{i(t-t_0)}{b^2}
\end{equation}
we can derive the following expressions\footnote{We did some additional $t$-independent shifts in $x$ and $y$ coordinates to adjust the coefficients in front of the exponential terms.}
\begin{multline}\label{sausage_UV_exp}
    ds^2_{\text{sausage}}=\frac{1}{2}\left(dx^2+dy^2\right)+ \\
    +e^{\frac{t-t_0}{b^2}}
    \left(e^{\frac{x}{b}}\left(dx+\frac{ib}{\sqrt{1+b^2}}dy\right)^2+e^{-\frac{x}{b}}\left(dx-\frac{ib}{\sqrt{1+b^2}}dy\right)^2\right)+\mathcal{O}\left(e^{\frac{3(t-t_0)}{b^2}}\right)
\end{multline}
and
\begin{multline}\label{lambda_UV_exp}
    ds^2_{\lambda}=\frac{1}{2}\left(dx^2+dy^2\right)+e^{\frac{t-t_0}{b^2}}
    \left(e^{\frac{x}{b}}\left(dx+\frac{ib}{\sqrt{1+b^2}}dy\right)^2+e^{-\frac{x}{b}}\left(dx-\frac{ib}{\sqrt{1+b^2}}dy\right)^2\right)+ \\
    +e^{\frac{t-t_0}{b^2}}
    e^{-\frac{iy}{\sqrt{1+b^2}}}\left(dx+\frac{ib}{\sqrt{1+b^2}}dy\right)\left(dx-\frac{ib}{\sqrt{1+b^2}}dy\right)+\mathcal{O}\left(e^{\frac{3(t-t_0)}{b^2}}\right)\,.
\end{multline}
An important observation is that the obtained UV expansions \eqref{sausage_UV_exp} and \eqref{lambda_UV_exp} are consistent with what we expect from the screening charges, which define the sausage and $\lambda$ models respectively (see appendix \ref{app_screenings}). This fact is an additional indication that  the obtained metrics \eqref{sausage-all-loop} and \eqref{lambda-deformed-metric-all-loop} provide correct completion of the corresponding one-loop theories at least up to the four loop order. 

We note also that the operator in the last line in \eqref{lambda_UV_exp} is irrelevant with the dimension $2+\frac{1}{1+b^2}$. However, making a $t$-dependent shift by an imaginary amount in $y$ direction, as we did in \eqref{b_def_shift}, one can make it formally small in the UV. Such manipulations are not well justified in QFT, however they shed some light on the meaning and role of $\lambda$-deformed sigma models. As YB-deformed sigma models correspond to asymptotically free theories, their $\lambda$-deformed cousins include some irrelevant operators. Note also that without the shift \eqref{b_def_shift} the UV limit of the metric \eqref{all_loop_lambda_metric_exp} is the non-trivial background
\begin{equation}\label{lambda_metric_rescaled_UV}
    ds^2=\frac{1}{2}\left(\frac{dx^2}{1-e^{-\frac{iy}{\sqrt{1+b^2}}}}+\frac{1-\frac{b^2}{1+b^2}e^{-\frac{iy}{\sqrt{1+b^2}}}}{\left(1-e^{-\frac{iy}{\sqrt{1+b^2}}}\right)^2}dy^2\right)\,,
\end{equation}
which solves the conformal equation
\begin{equation}
    \beta_{ij}(G)+2\nabla_i\nabla_j\Phi=0\quad\text{with}\quad \Phi=-\frac{1}{2}\log\left(1-e^{\frac{iy}{\sqrt{1+b^2}}}\right)-\frac{1}{4(1+b^2)\left(1-e^{\frac{iy}{\sqrt{1+b^2}}}\right)}
\end{equation}
with the ``exact'' $\beta$-function \eqref{all-loop-beta-function-concl}.

The metric \eqref{lambda_metric_rescaled_UV} corresponds to  $SU(2)/U(1)$ gauged WZW model. It is more convenient to consider the related $SL(2,\mathbb{R})/U(1)$ metric which is obtained from  \eqref{lambda_metric_rescaled_UV} by replacing $x\leftrightarrow y$ and $i\sqrt{1+b^2}\leftrightarrow b$
\begin{equation}\label{lambda_metric_rescaled_UV-2}
    ds^2=\frac{1}{2}\left(\frac{1-\frac{1+b^2}{b^2}e^{\frac{x}{b}}}{\left(1-e^{\frac{x}{b}}\right)^2}dx^2+\frac{dy^2}{1-e^{\frac{x}{b}}}\right)\,.
\end{equation}
Changing the variables
\begin{equation}
    x=b\log\left(1-\coth^2r\right),\quad y=b\varphi
\end{equation}
and then replacing $b^2=\frac{1}{\hbar}$ one finds that \eqref{lambda_metric_rescaled_UV-2} reduces to
\begin{equation}\label{our-cigar}
    ds^2=\frac{2}{\hbar}\left(\Big(1+\frac{\hbar}{\cosh^2r}\Big)dr^2+\tanh^2rd\varphi^2\right).
\end{equation}
This metric corresponds to all-loop completion of the cigar metric \cite{Witten:1991yr} in our scheme.

In \cite{Dijkgraaf:1991ba} Dijkgraaf, Verlinde and Verlinde suggested the following expression for the exact cigar metric (see also \cite{Bars:1992sr})
\begin{equation}\label{DVV-cigar}
    ds^2_{\text{DVV}}=\frac{2}{\hbar}\left(dr^2+\frac{1}{\coth^2r-\frac{\hbar}{\hbar+1}}d\varphi^2\right)
\end{equation}
Two metrics \eqref{our-cigar} and \eqref{DVV-cigar} should be related by some covariant metric redefinition \eqref{cov_red_metric} together with a coordinate change
\begin{equation}
    r\rightarrow r+\hbar f_1(r)+\hbar^2 f_2(r)+\dots,\quad \varphi\rightarrow (1+\hbar q_1+\hbar^2 q_2+\dots)\varphi
\end{equation}
One finds that this covariant metric redefinition is very simple
\begin{equation}
    \left(1-R\right)^{\frac{1}{2}}ds^2=ds^2_{\text{DVV}}\left(\varphi\rightarrow\frac{\varphi}{\sqrt{1+\hbar}}\right)=\frac{2}{\hbar}\left(dr^2+\frac{\frac{1}{1+\hbar}}{\coth^2r-\frac{\hbar}{\hbar+1}}d\varphi^2\right)
\end{equation}
i.e. it corresponds to the following choice of the scheme parameters in \eqref{cov_red_metric}
\begin{equation}
    c_1=-\frac{1}{2},\quad c_2=-\frac{1}{8},\quad c_5=-\frac{1}{16},\quad c_3=c_4=c_6=c_7=c_8=c_9=c_{10}=c_{11}=0\,.
\end{equation}

We note that applying the same scheme transformation $ds^2\rightarrow(1-R)^{\frac{1}{2}}ds^2$ to \eqref{sausage-all-loop} gives the metric
\begin{equation}\label{extra-factor}
    ds^2=2\kappa
    \left(\big(\kappa-\hbar^{-1}\big)\big(\kappa^{-1}-\hbar^{-1}\big)\right)^{\frac{1}{2}}
    \left(\frac{d\theta^2}{1-\kappa^2\sin^2\theta}+\frac{\cos^2 \theta d\chi^2}{1-\kappa\big(\kappa\sin^2\theta+\hbar\cos^2\theta\big)}\right)
\end{equation}
which coincides with the metric from \cite{Hoare:2019ark} (equation (2.6) in \cite{Hoare:2019ark}  with $\varkappa=i\kappa$, $r=\sin\theta$ and $h=\frac{2\kappa}{\hbar}$) by an overall factor
\begin{equation}
    \frac{2\kappa
    \left(\big(\kappa-\hbar^{-1}\big)\big(\kappa^{-1}-\hbar^{-1}\big)\right)^{\frac{1}{2}}}{\frac{2\kappa}{\hbar}-1-\kappa^2}=1-\frac{(\kappa^2-1)^2}{8\kappa^2}\hbar^2-\frac{(\kappa^2-1)^2(\kappa^2+1)}{8\kappa^3}\hbar^3+O(\hbar^{4})
\end{equation}
We see that these two metrics coincide at two loops, but start to differ at the third one. However, the extra factor \eqref{extra-factor} can be interpreted as a particular type of  metric redefinitions (in general non-covariant) of the form
\begin{equation}
    G_{ij}\rightarrow\left(1+\xi_1(t)\hbar+\xi_2(t)\hbar^2+\dots\right)G_{ij}\,,
\end{equation}
which implies that both metrics are directly related.
\section{Concluding remarks}\label{concl}

In this paper we studied the question of renormalizability of $2D$ integrable sigma models. Namely, we considered two integrable backgrounds, both corresponding to the deformations of the $O(3)$ model, known as YB-deformed and $\lambda$-deformed $SU(2)/U(1)$ sigma models. We came to the conclusion similar to already posted in \cite{Hoare:2019ark,Hoare:2019mcc}, that higher loop corrections to the classical metrics are needed to ensure renormalizability. We have found the scheme with the $\beta$-function \eqref{all-loop-beta-function-first-4-loops} in which the metric receives only $\hbar^0$ correction (the metric is two-loop exact). It is important that both backgrounds, namely  \eqref{sausage-all-loop} and \eqref{lambda-deformed-metric-all-loop}, satisfy additional constraint \eqref{special-condition} making $(2+\zeta_3)$ term in the fourth loop  coefficient of the $\beta-$function negligible. Remarkably both solutions  \eqref{sausage-all-loop} and \eqref{lambda-deformed-metric-all-loop} also solve RG flow equation with the ``exact'' $\beta$-function \eqref{all-loop-beta-function-concl}, which we obtained by naive extrapolation of the four-loop result \eqref{all-loop-beta-function-first-4-loops}.

It is interesting to extend our study for sigma models with the dimension of the target space $D>2$. In general, this will require the knowledge of $\beta$-function including the one for the Kalb-Ramond field ($B$-field), which appears in most of deformed integrable sigma models. It makes the analysis more problematic since the $\beta$-function in the presence of the $B$-field is known only in two-loop approximation. However, there are models, such as $\lambda$-deformed $SO(N+1)/SO(N)$ model, where the $B$-field vanishes \cite{Grigoriev:2007bu,Demulder:2015lva}. We plan to consider the renormalizability of these models in a separate publication.

Another possible direction of development of our results is to study the $\eta$-deformed $OSp(N|2m)$ sigma models, which contain anticommuting coordinates in their target space. The RG flow of these models depends on the difference of $N-2m$. It was shown in \cite{Alfimov:2020jpy} that in the regime of purely imaginary $\eta=i\kappa$ they are asymptotically free if $N-2m>2$, conformal if $N-2m=2$ and infrared free if $N-2m<2$. There was also analyzed the UV limit of these theories for $N-2m>2$, which can be understood as a free theory perturbed by a set of operators assembled from certain screening charges. Interestingly, the 1-loop RG flow equation for asymptotically free deformed $OSP(5|2)$ sigma model leads to the same differential equation for $\kappa$ as \eqref{kappa-one-loop-flow-equation} for the deformed $O(3)$ sigma-model. This similarity to the $O(3)$ sigma model RG flow together with the fact that we know the UV limit of the $OSp(5|2)$ action from the screenings gives some hope to find a simple expression like \eqref{all-loop-beta-function-concl} for the $\beta$-function, despite the target space of the $OSp(5|2)$ has the dimension larger than $2$.
\section*{Acknowledgments}
We are grateful to Boris Feigin, Ben Hoare, Sergei Lukyanov, Konstantinos Sfetsos, Konstantinos Siampos, Arkady Tseytlin, Mikhail Vasiliev and Alexander Zamolodchikov for fruitful and stimulating discussions. A.L. acknowledges the support of Basis Foundation. M.A. would like to thank Elena M. for her good advice at the initial stage of this project.
\appendix
\section{\texorpdfstring{$\beta$}{beta}-function in different schemes}\label{schemes}
In the present Appendix we address the question how the $\beta$-function in the $D=2$ case changes under covariant metric redefinitions starting from the minimal subtraction scheme. Let us describe the way to find the contributions to the transformed $\beta$-function order by order in the covariant expansion, using the Mathematica package xAct \cite{Martin:xAct,Brizuela:2008ra,Nutma:2013zea} for the necessary calculations.

Here we denote by tilde the metric after covariant redefinition, namely
\begin{equation}\label{cov_met_redef}
    \tilde{G}_{ij}=G_{ij}+\sum_{k=0}^{\infty}G_{ij}^{(k)}\,,
\end{equation}
where $G_{ij}^{(k)}$ are given by \eqref{cov_red_metric}. Essentially we are to determine the $\tilde{b}_k^{(L)}$ coefficients in front of the tensor structures from \eqref{beta_function_tensor_exp} in the new scheme. For the case of two-dimensional target space the number of independent tensor structures is much lower and we can explicitly write them in the first four orders we are interested in
\begin{align}
    \mathcal{T}_1 &= \left\{RG_{ij}\right\}\;, \quad \mathcal{T}_2=\left\{R^2 G_{ij}, \nabla^2 RG_{ij},  \nabla_i \nabla_j R\right\}\,, \notag \\
    \mathcal{T}_3 &= \left\{R^3 G_{ij}, R\nabla^2 RG_{ij}, (\nabla R)^2 G_{ij}, \nabla^2 \nabla^2 RG_{ij}, \nabla_i R\nabla_j R\right\}\,, \notag \\
    \mathcal{T}_4 &= \left\{R^4 G_{ij}, R^2 \nabla^2 RG_{ij}, R (\nabla R)^2 G_{ij}, R\nabla^2 \nabla^2 RG_{ij}, (\nabla^2 R)^2 G_{ij}, (\nabla_k \nabla_l R)(\nabla^k \nabla^l R)G_{ij}\,, \right. \notag \\
    & \left. \nabla_k \nabla^2 R \nabla^k RG_{ij}, \nabla^2 \nabla^2 \nabla^2 RG_{ij}\;, R\nabla_i R\nabla_j R\,, \nabla^2 R \nabla_i \nabla_j R\right\}\,.
\end{align}
One may notice the absence of some possible structures in the space $\mathcal{T}_3$, such as $R\nabla_i \nabla_j R$ and $\nabla_i \nabla_j \nabla^2 R$. The reason for their exclusion is that the sum of the first of them with $\nabla_i R\nabla_j R$ is of the form $\nabla_i V_j+\nabla_j V_i$, i.e. can be absorbed by a diffeomorphism, and the second one has already the form of a diffeomorphism. For the same reason we exclude from the space $\mathcal{T}_4$ the structures $R^2 \nabla_i \nabla_j R$, $\nabla_i R \nabla_j \nabla^2 R$, $\nabla_j R \nabla_i \nabla^2 R$, where the latter two can be simultaneously eliminated due to the symmetry $i \leftrightarrow j$, $R \nabla_i \nabla_j \nabla^2 R$ and $\nabla_k R \nabla_i \nabla_j \nabla^k R$.

In the 4th order there is the following identity, which also diminishes the number of possible tensor structures in $\mathcal{T}_4$
\begin{equation}\label{ident}
    \nabla_i \nabla_k R \nabla_j \nabla^k R=\nabla^2 R \nabla_i \nabla_j R-\frac{1}{2}(\nabla^2 R)^2 G_{ij}+\frac{1}{2}(\nabla_k \nabla_l R)(\nabla^k \nabla^l R)G_{ij}\,.
\end{equation}
Let us first rewrite the terms proportional to the metric in the following form, utilizing the expression for Ricci tensor in 2D
\begin{equation}\label{prop_to_metric}
    \frac{1}{2}(\nabla_k \nabla_l R)(\nabla^k \nabla^l R)G_{ij}-\frac{1}{2}(\nabla^2 R)^2 G_{ij}=\nabla^k \nabla^l R \left(\nabla_k \nabla_l R_{ij}-\frac{1}{2}\nabla^2 (RG_{kl} G_{ij})\right)\,.
\end{equation}
For the first term in the brackets on the right hand side of \eqref{prop_to_metric} we use the second contracted Bianchi identity
\begin{equation}\label{ident1}
    \nabla_k \nabla_l R_{ij}=\nabla_k \nabla_i R_{lj}-\nabla_k \nabla^m R_{lijm}\,.
\end{equation}
For the second term we obtain
\begin{equation}\label{ident2}
    \frac{1}{2}\nabla^m \nabla_m (RG_{kl} G_{ij})=\nabla^m \nabla_m R_{kjli}+\frac{1}{2}\nabla^m \nabla_m (RG_{ki} G_{lj})\;.
\end{equation}
By substituting \eqref{ident1} and \eqref{ident2} into \eqref{prop_to_metric}, we derive, using the fact that covariant derivatives commute on Riemann tensor in 2D and general symmetry properties of the Riemann tensor together with the second Bianchi identity,
\begin{align}
    &\frac{1}{2}\nabla_i \nabla_k R\nabla_i \nabla^k R-\frac{1}{2}\nabla_k \nabla^k R\nabla_i \nabla_j R-\nabla^k \nabla^l R\nabla^m (\nabla_k R_{lijm}+\nabla_m R_{kjli})= \notag \\
    =&\frac{1}{2}\nabla_i \nabla_k R\nabla_j \nabla^l R-\frac{1}{2}\nabla_k \nabla^k R\nabla_i \nabla_j R+\nabla^k \nabla^l R\nabla^m \nabla_j R_{limk}= \notag \\
    =&\nabla_i \nabla_k R\nabla_j \nabla^k R-\nabla_k \nabla^k R\nabla_i \nabla_j R\,.
\end{align}
Thus, we have proven the initial identity \eqref{ident}. This identity allows to eliminate the tensor structure $\nabla_i \nabla_k R \nabla_j \nabla^k R$ from $\mathcal{T}_4$.

In addition we do not include the structures which differ by the permutations of the covariant derivatives acting on the same scalar. In this way we exclude the structures such as all the permutations of the derivatives in $\nabla^2 \nabla^2 RG_{ij}$ in $\mathcal{T}_3$ and $\nabla^2 \nabla^2 \nabla^2 RG_{ij}$ in $\mathcal{T}_4$ and $\nabla^2 \nabla_k R \nabla^k RG_{ij}$, $\nabla_i R \nabla^2 \nabla_j R$, $\nabla_j R \nabla^2 \nabla_i R$, $\nabla^k R\nabla_j \nabla_i \nabla_k R$ and $\nabla^k R\nabla_k \nabla_i \nabla_j R$ also in $\mathcal{T}_4$, because they are linearly dependent with the remaining ones. Thus, we are able to write the expansion \eqref{beta_function_tensor_exp} in the first four orders explicitly and find the coefficients $\tilde{b}_k^{(L)}$, where $k=1,\ldots,\textrm{dim}\,\mathcal{T}_L$ in the new scheme.

Let us now describe how the $\beta$-function changes under covariant substitution of the metric. In the new scheme we have in general
\begin{equation}\label{RG_eq_new_scheme}
    \dot{\tilde{G}}_{ij}=-(\tilde{\beta}_{ij}+\tilde{\nabla}_i \tilde{V}_j+\tilde{\nabla}_j \tilde{V}_i)(\tilde{G}_{mn})\,.
\end{equation}
We substitute \eqref{cov_met_redef} to \eqref{RG_eq_new_scheme} to obtain
\begin{equation}\label{RG_eq_new_scheme_covariant_subst}
    \dot{G}_{ij}+\sum\limits_{k=0}^{+\infty}\dot{G}_{ij}^{(k)}=-(\tilde{\beta}_{ij}+\tilde{\nabla}_i \tilde{V}_j+\tilde{\nabla}_j \tilde{V}_i)\left(G_{mn}+\sum\limits_{k=0}^{+\infty}G_{mn}^{(k)}\right)\,.
\end{equation}
As every $G_{ij}^{(k)}$ can be expressed solely in terms of $G_{ij}$ itself, the left hand side of \eqref{RG_eq_new_scheme_covariant_subst} is transformed to
\begin{equation}\label{K_operators}
    \dot{G}_{ij}+\sum\limits_{k=0}^{+\infty}\dot{G}_{ij}^{(k)}=\left(\delta_i^l+\sum\limits_{k=1}^{+\infty}\left(K_i^l\right)^{(k)}\right)\dot{G}_{lj}\,,
\end{equation}
in which each operator $(K_i^l)^{(k)}$ is of order $\hbar^k$ and is a function of the coefficients $c_p$, $p=1,\ldots,+\infty$ from \eqref{cov_red_metric} and $G_{ij}$ itself, possibly acting on $\dot{G}_{lj}$ with covariant derivatives. Combining \eqref{K_operators} with \eqref{RG_eq_new_scheme_covariant_subst}, we can extract the RG flow equation
\begin{equation}\label{RG-flow-equation-cr}
    \dot{G}_{ij}=- (\tilde{\beta}_{ij}+\tilde{\nabla}_i \tilde{V}_j+\tilde{\nabla}_j \tilde{V}_i)\left(G_{mn}+\sum\limits_{k=0}^{+\infty}G_{mn}^{(k)}\right)-\sum\limits_{k=1}^{+\infty}\left(K_i^l\right)^{(k)}\dot{G}_{lj}\,,
\end{equation}
which should be identical to \eqref{RG-flow-equation} after substituting $\dot{G}_{ij}$ in the minimal subtraction scheme to the right hand side. Therefore, the difference between the right hand side of \eqref{RG-flow-equation-cr} and \eqref{RG-flow-equation} has to be $0$, which leads to the set of equations for the coefficients $\tilde{b}_k^{(L)}$, $b_k^{(L)}$ and $c_p$ order by order in $\hbar$
\begin{equation}\label{scheme_change_eq}
 (\tilde{\beta}_{ij}+\tilde{\nabla}_i \tilde{V}_j+\tilde{\nabla}_j \tilde{V}_i)\left(G_{mn}+\sum\limits_{k=0}^{+\infty}G_{mn}^{(k)}\right)-\beta_{ij}(G_{mn})=0\,.
\end{equation}

Let us track how the $\beta$-function changes in the first 3 orders and then present the result for the 4th order. The expressions at these orders in the minimal subtraction scheme were written in \eqref{beta-123} and \eqref{beta-4}, while in new scheme they are given by the ansatz (1-loop $\beta$-function is omitted as it is trivially non-dependent on the scheme choice)
\begin{align}
    \tilde{\beta}^{(2)}_{ij}&=\left(\tilde{b}_1^{(2)}\tilde{R}^2+\tilde{b}_2^{(2)}\tilde{\nabla}^2 \tilde{R}\right)\tilde{G}_{ij}+\tilde{b}_3^{(2)}\tilde{\nabla}_i \tilde{\nabla}_j \tilde{R}\,, \notag \\
    \tilde{\beta}^{(3)}_{ij}&=\left(\tilde{b}_1^{(3)}\tilde{R}^3+\tilde{b}_2^{(3)}\tilde{R}\tilde{\nabla}^2 \tilde{R}+\tilde{b}_3^{(3)}(\tilde{\nabla} \tilde{R})^2+\tilde{b}_4^{(3)}\tilde{\nabla}^2 \tilde{\nabla}^2 \tilde{R}\right)\tilde{G}_{ij}+\tilde{b}_5^{(3)}\tilde{\nabla}_i \tilde{R}\tilde{\nabla}_j \tilde{R}\,, \notag \\
    \tilde{\beta}^{(4)}_{ij}&=\left(\tilde{b}_1^{(4)}\tilde{R}^4+\tilde{b}_2^{(4)}\tilde{R}^2 \tilde{\nabla}^2 \tilde{R}+\tilde{b}_3^{(4)}\tilde{R}(\tilde{\nabla} \tilde{R})^2+\tilde{b}_4^{(4)}\tilde{R}\tilde{\nabla}^2 \tilde{\nabla}^2 \tilde{R}+\tilde{b}_5^{(4)}(\tilde{\nabla}^2 \tilde{R})^2+\right. \notag \\
    &\left.+\tilde{b}_6^{(4)}(\tilde{\nabla}_k \tilde{\nabla}_l \tilde{R})(\tilde{\nabla}^k \tilde{\nabla}^l R)+\tilde{b}_7^{(4)}(\tilde{\nabla}_k \tilde{\nabla}^2 \tilde{R})(\tilde{\nabla}^k \tilde{R})+\tilde{b}_8^{(4)}\tilde{\nabla}^2 \tilde{\nabla}^2 \tilde{\nabla}^2 \tilde{R}\right)\tilde{G}_{ij}+ \notag \\
    &+\tilde{b}_9^{(4)}\tilde{R}\tilde{\nabla}_i \tilde{R}\tilde{\nabla}_j \tilde{R}+\tilde{b}_{10}^{(4)}\tilde{\nabla}^2 \tilde{R} \tilde{\nabla}_i \tilde{\nabla}_j \tilde{R}\,, \label{beta_ansatz_new_scheme}
\end{align}
where tilde over $\tilde{\nabla}$ and $\tilde{R}$ means that they are calculated with respect to the metric $\tilde{G}_{ij}$.
Putting \eqref{beta_ansatz_new_scheme}, \eqref{beta-123}, \eqref{beta-4} and \eqref{cov_red_metric} into \eqref{scheme_change_eq}, we derive a set of equations for the unknown coefficients $\tilde{b}_k^{(L)}$. These equations are too cumbersome to be written here, but can be easily solved and here we write just the solution for the two- and three-loop orders
\begin{align}
    \tilde{b}_1^{(2)}&=\frac{1}{4}\,, \quad \tilde{b}_2^{(2)}=0\,, \quad \tilde{b}_3^{(2)}=0\,, \notag \\
    \tilde{b}_1^{(3)}&=\frac{5}{32}+\frac{c_1}{4}-\frac{c_2}{2}\,, \quad \tilde{b}_2^{(3)}=\frac{1}{16}-\frac{c_1}{2}+c_2-c_3-\frac{c_1^2}{2}\,, \quad \tilde{b}_3^{(3)}=-c_3-\frac{c_1^2}{2}\,, \notag \\
    \tilde{b}_4^{(3)}&=0\,, \quad \tilde{b}_5^{(3)}=-\frac{1}{16}
\end{align}
and for the four-loop order
\begin{align}
    \tilde{b}_1^{(4)}&=\frac{23}{192}+\frac{3c_1}{8}-c_5-\frac{c_1^2}{4}\,, \quad \tilde{b}_2^{(4)}=\frac{2+\zeta_3}{32}+\frac{3c_1}{32}-\frac{c_3}{2}-\frac{3c_7}{2}-c_1^2-\frac{3c_1 c_2}{2}+\frac{c_1^3}{2}\,, \notag \\
    \tilde{b}_3^{(4)}&=\frac{41+12\zeta(3)}{192}-\frac{15c_1}{16}-2c_3+3c_5-\frac{3c_6}{2}-c_7-c_8+\frac{c_9}{2}-\frac{c_{10}}{2}-\frac{c_1^2}{2}-3c_1 c_2+ \notag \\
    &+\frac{5c_1 c_4}{4}+c_1^3\,, \notag \\
    \tilde{b}_4^{(4)}&=-\frac{3c_8}{2}-\frac{c_1 c_3}{2}\,, \notag \\
    \tilde{b}_5^{(4)}&=\frac{1}{192}-\frac{c_1}{16}-\frac{c_3}{2}-\frac{3c_8}{2}-\frac{c_9}{2}+\frac{c_{10}}{2}-\frac{c_1^2}{4}-\frac{c_1 c_3}{2}-\frac{c_1 c_4}{2}\,, \notag \\
    \tilde{b}_6^{(4)}&=\frac{1}{192}-\frac{c_1}{16}-c_3+c_6-2c_8+\frac{c_9}{2}-\frac{c_{10}}{2}\,, \notag \\
    \tilde{b}_7^{(4)}&=-\frac{c_1}{8}-2c_3+c_7-5c_8-c_1 c_3\,, \quad \tilde{b}_8^{(4)}=0\,, \notag \\
    \tilde{b}_9^{(4)}&=-\frac{25}{192}-\frac{5c_1}{16}-\frac{5c_9}{2}+\frac{5c_{10}}{2}-\frac{5c_1 c_4}{2}\,, \quad \tilde{b}_{10}^{(4)}=-\frac{1}{96}+\frac{c_1}{8}+c_9-c_{10}+3c_1 c_4\,.
\end{align}

If we want to transform to the $\beta$-function \eqref{all-loop-beta-function-first-4-loops}, then the scheme parameters $c_k$ have to be chosen as
\begin{align}
    c_2 &= -\frac{1}{16}+\frac{c_1}{2}\,, & c_3 &= -\frac{c_1^2}{2}\,, \notag \\
    c_5 &= -\frac{7}{192}+\frac{3c_1}{8}-\frac{c_1^2}{4}\,, & c_6 &= \frac{9+5\zeta_3}{96}+\frac{c_1}{8}-\frac{c_1^2}{2}+\frac{c_1 c_4}{2}+\frac{c_1^3}{3}\,, \notag \\
    c_7 &= \frac{5(2+\zeta_3)}{96}+\frac{c_1}{8}-c_1^2+\frac{c_1^3}{3}\,, & c_8 &= \frac{2+\zeta_3}{96}+\frac{c_1^3}{6}\,, \notag \\
    c_9-c_{10} &= \frac{1}{96}-\frac{c_1}{8}-c_1 c_4\,. \label{solution_c_coefficients}
\end{align}
Given the choice \eqref{solution_c_coefficients}, we arrive to particularly simple expressions for the $\beta$-function in the first four orders
\begin{align}
    \tilde{\beta}^{(1)}_{ij} &= \frac{1}{2}\tilde{R}\tilde{G}_{ij}\,, \quad \tilde{\beta}^{(2)}_{ij} = \frac{1}{4}\tilde{R}^2 \tilde{G}_{ij}\,, \quad \tilde{\beta}^{(3)}_{ij} = \frac{3}{16}\tilde{R}^3 \tilde{G}_{ij}-\frac{1}{16}\tilde{\nabla}_i \tilde{R}\tilde{\nabla}_j \tilde{R}\,, \notag \\
    \tilde{\beta}^{(4)}_{ij} &= \frac{5}{32}\tilde{R}^4 \tilde{G}_{ij}-\frac{5}{32}\tilde{R}\tilde{\nabla}_i \tilde{R}\tilde{\nabla}_j \tilde{R}+\frac{2+\zeta_3}{64}\tilde{\nabla}^2 \left(\tilde{R}^3+2\tilde{R}\tilde{\nabla}^2 \tilde{R}-\frac{1}{2}\tilde{\nabla}^2 \tilde{R}^2 \right)\,.
\end{align}
\section{Deformed \texorpdfstring{$O(3)$}{O(3)} sigma model screening charges}\label{app_screenings}

In the work \cite{Litvinov:2018bou} it was shown that integrable deformed $O(N)$ sigma models in the UV limit can be understood as a Gaussian theory perturbed by certain relevant operators (for the $OSp(N|2m)$ sigma models see the discussion in \cite{Alfimov:2020jpy}). These operators are constrained by the integrable structure of the sigma model and are built from the screening charges. The screening charges determine the integrals of motion of the sigma model and vice versa (see \cite{Litvinov:2018bou}).

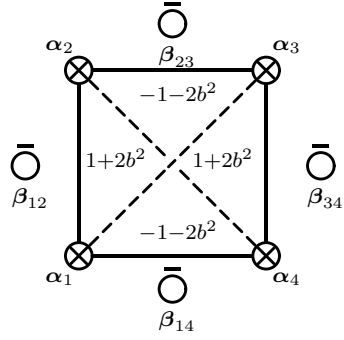
\begin{figure}[t]
\begin{picture}(200,140)(0,60)
\Thicklines
\unitlength 5pt
\put(37,32){\circle{2}}
\put(37,18){\circle{2}}
\put(36.4,31.4){\line(1,1){1.2}}
\put(36.4,32.6){\line(1,-1){1.2}}
\put(36.4,17.4){\line(1,1){1.2}}
\put(36.4,18.6){\line(1,-1){1.2}}
\put(37,19){\line(0,1){12}}
\put(44.2,25.2){\line(-1,-1){0.5}}
\put(43.4,24.4){\line(-1,-1){0.6}}
\put(42.4,23.4){\line(-1,-1){0.6}}
\put(41.4,22.4){\line(-1,-1){0.6}}
\put(40.4,21.4){\line(-1,-1){0.6}}
\put(39.4,20.4){\line(-1,-1){0.6}}
\put(38.4,19.4){\line(-1,-1){0.6}}
\put(43.4,25.6){\line(-1,1){0.6}}
\put(42.4,26.6){\line(-1,1){0.6}}
\put(41.4,27.6){\line(-1,1){0.6}}
\put(40.4,28.6){\line(-1,1){0.6}}
\put(39.4,29.6){\line(-1,1){0.6}}
\put(38.4,30.6){\line(-1,1){0.6}}
\put(33,24.8){\circle{2}}
\put(32.4,26.3){\line(1,0){1.2}}
\put(32,22){$\scriptstyle{\boldsymbol{\beta}_{12}}$}
\put(37.5,24.7){$\scriptstyle{1+2b^{2}}$}
\put(45.5,24.7){$\scriptstyle{1+2b^{2}}$}
\put(41.5,29.7){$\scriptstyle{-1-2b^{2}}$}
\put(41.5,19.3){$\scriptstyle{-1-2b^{2}}$}
\put(55.1,24.8){\circle{2}}
\put(54.5,26.3){\line(1,0){1.2}}
\put(54.1,22){$\scriptstyle{\boldsymbol{\beta}_{34}}$}
\put(34.5,16){$\scriptstyle{\boldsymbol{\alpha}_{1}}$}
\put(34.5,33.5){$\scriptstyle{\boldsymbol{\alpha}_{2}}$}
\put(51.5,33.5){$\scriptstyle{\boldsymbol{\alpha}_{3}}$}
\put(51.5,16){$\scriptstyle{\boldsymbol{\alpha}_{4}}$}
\put(44,35.5){\circle{2}}
\put(43.4,37){\line(1,0){1.2}}
\put(43,32.7){$\scriptstyle{\boldsymbol{\beta}_{23}}$}
\put(44,15.5){\circle{2}}
\put(43.4,17){\line(1,0){1.2}}
\put(43,12.7){$\scriptstyle{\boldsymbol{\beta}_{14}}$}
\put(51,32){\circle{2}}
\put(51,18){\circle{2}}
\put(50.4,31.4){\line(1,1){1.2}}
\put(50.4,32.6){\line(1,-1){1.2}}
\put(50.4,17.4){\line(1,1){1.2}}
\put(50.4,18.6){\line(1,-1){1.2}}
\put(51,19){\line(0,1){12}}
\put(44.6,24,4){\line(1,-1){0.6}}
\put(45.6,23,4){\line(1,-1){0.6}}
\put(46.6,22,4){\line(1,-1){0.6}}
\put(47.6,21,4){\line(1,-1){0.6}}
\put(48.6,20,4){\line(1,-1){0.6}}
\put(49.6,19,4){\line(1,-1){0.6}}
\put(44.6,25,6){\line(1,1){0.6}}
\put(45.6,26,6){\line(1,1){0.6}}
\put(46.6,27,6){\line(1,1){0.6}}
\put(47.6,28,6){\line(1,1){0.6}}
\put(48.6,29,6){\line(1,1){0.6}}
\put(49.6,30,6){\line(1,1){0.6}}
\put(38,18){\line(1,0){12}}
\put(38,32){\line(1,0){12}}
\end{picture}
\caption{Screening charges corresponding to the deformed $O(3)$ sigma model.}
\label{O3_screenings}
\end{figure}

The deformed $O(3)$ sigma model has two bosonic degrees of freedom and therefore can be encoded as a two component real bosonic field. Let us recall that the theory in question may be determined by the set of screening fields on the figure \ref{O3_screenings} (see for example \cite{Fateev:2017mug,Litvinov:2018bou}),
where the crosses correspond to the fermionic screening fields
\begin{equation}\label{fermionic_screening}
\mathcal{S}_k=\oint e^{(\boldsymbol{\alpha}_k, \varphi)}dz,
\end{equation}
which define the dual Toda-like description of the ``sausage'' model, while the circles with the bar to the so called dressed or Wakimoto ones
\begin{equation}\label{dressed_screening}
    \mathcal{S}_{ij}=\oint e^{(\boldsymbol{\beta}_{ij}\cdot\boldsymbol{\varphi})}
    \big(\boldsymbol{\alpha}_i\cdot\partial\boldsymbol{\varphi}\big)dz\,,
\end{equation}
which comprise the relevant perturbation operators in the UV limit of the sigma model and where $\boldsymbol{\varphi}$ in our case denotes the pair of chiral real bosonic fields. By utilizing Cartesian coordinates as in \cite{Litvinov:2018bou} we can parametrize the fermionic screening lengths as follows
\begin{align}\label{fermionic_screening_lengths}
    & \boldsymbol{\alpha}_1 =bE_1+i\sqrt{1+b^2} e_1\,, \quad \boldsymbol{\alpha}_2 =bE_1-i\sqrt{1+b^2} e_1\,, \notag \\
    & \boldsymbol{\alpha}_3 =-bE_1+i\sqrt{1+b^2} e_1\,, \quad \boldsymbol{\alpha}_4 =-bE_1-i\sqrt{1+b^2} e_1\,,
\end{align}
where the parameter $b$ is the same as in the formula \eqref{b_def_shift} and $\beta_{ij}$ lie in the same linear space spanned by $E_1$ and $e_1$. The values on the links on the figure \ref{O3_screenings} correspond to the scalar products of the $\alpha_i$'s on its ends, being $1$ for the dashed ones. In addition, there was shown in \cite{Litvinov:2016mgi} that each pair of fermionic screenings $\mathcal{S}_i$ and $\mathcal{S}_j$ of the type \eqref{fermionic_screening} can be equivalently replaced by the  dressed or Wakimoto screening $\mathcal{S}_{ij}$ of the type \eqref{dressed_screening}, whose length is determined by the formula
\begin{equation}
    \boldsymbol{\beta}_{ij}=\frac{2(\boldsymbol{\alpha}_i+\boldsymbol{\alpha}_j)}{(\boldsymbol{\alpha}_i+\boldsymbol{\alpha}_j)^2}
\end{equation}
and in the deformed $O(3)$ case they are given by
\begin{equation}\label{dressed_screening_lengths}
    \boldsymbol{\beta}_{12}=\frac{E_1}{b}\,, \quad \boldsymbol{\beta}_{34}=-\frac{E_1}{b}\,, \quad \boldsymbol{\beta}_{13}=-i\frac{e_1}{\sqrt{1+b^2}}\,, \quad \boldsymbol{\beta}_{24}=i\frac{e_1}{\sqrt{1+b^2}}\,.
\end{equation}
An important feature of the dressed screenings is that in their prefactor we can always shift $\alpha_i \rightarrow \alpha_i+\xi \beta_{ij}$, because this shift is proportional to total derivative. All this allows us to say that in the sausage metric in the UV limit \eqref{sausage_UV_exp} we observe the screenings $\mathcal{S}_{12}$ and $\mathcal{S}_{34}$. Namely, if we take $\xi=1$, then the screenings $\mathcal{S}_{12}$ and $\mathcal{S}_{34}$ turn to
\begin{equation}
    \mathcal{S}_{12}=\oint dz b\left(\partial\boldsymbol{\varphi}_1+\frac{ib\partial\boldsymbol{\varphi}_2}{\sqrt{1+b^2}}\right)e^{\frac{\boldsymbol{\varphi}_1}{b}}\,, \quad \mathcal{S}_{34}=\oint dz b\left(\partial\boldsymbol{\varphi}_1-\frac{ib\partial\boldsymbol{\varphi}_2}{\sqrt{1+b^2}}\right)e^{-\frac{\boldsymbol{\varphi}_1}{b}}\,,
\end{equation}
which, after being supplemented by the antichiral part and identifying $\boldsymbol{\varphi}_1$ with $x$ and $\boldsymbol{\varphi}_2$ with $y$, reproduce the exponential terms in \eqref{sausage_metric_rescaled}.

The $\lambda$ model does not flow to the free theory in the UV limit, thus we can expect that in this limit it is given by a free theory perturbed by some irrelevant operator. And it is indeed the case: the screening $\mathcal{S}_{13}$ taken with the different shift parameter $\xi=-1$
\begin{equation}
    \mathcal{S}_{13}=\oint dz b\left(\partial\boldsymbol{\varphi}_1-\frac{ib\partial\boldsymbol{\varphi}_2}{\sqrt{1+b^2}}\right)e^{-\frac{i\boldsymbol{\varphi}_1}{\sqrt{1+b^2}}}\,,
\end{equation}
after multiplication by the antichiral part and the same identification of $\boldsymbol{\varphi}$ yields the third exponential term in \eqref{lambda_metric_rescaled}. Therefore, the difference between \eqref{sausage_metric_rescaled} and \eqref{lambda_metric_rescaled} is that in the latter there is the irrelevant perturbation term, which makes the corresponding theory not free in the UV limit.

\bibliographystyle{JHEP}
\bibliography{MyBib}

\end{document}